# The generation of gravitational waves

A. Loinger *

*Dipartimento di Fisica, Università di Milano, Via Celoria, 16 - 20133 Milano, Italy*

Abstract. – A proof that the generation of gravitational waves is physically impossible.

Many papers have been written on the gravitational waves, see e.g. the ample bibliography at the end of the review article by Schutz [1]. Now, I have recently demonstrated the non-existence of a *physical* "mechanism" for the production of gravitational waves [2]. In the present Letter I give a simpler proof of this thesis.

As is known, we can *always* choose a Gaussian ("synchronous", in Landau's terminology) system of space-time coordinates [3]. In such a reference frame the four-dimensional interval has the following form:

$$\mathrm{d}s^2 = c^2 \mathrm{d}t^2 - g_{\alpha\beta}(x^1, x^2, x^3, t)\mathrm{d}x^\alpha \mathrm{d}x^\beta \ , \ (\alpha, \beta = 1,2,3) \ , \tag{1}$$

and the time lines *coincide* with the geodesics.

Let us consider several point masses which interact *via* gravitational forces only. Of course, their motions can be quite complicated, but we are sure that any particle follows a geodesic line. Now, a geodesic motion is a ***free*** motion, it is the perfect physical analogue of a rectilinear and uniform motion of a pointlike electric charge in the customary Maxwell-Lorentz electrodynamics. No electromagnetic wave is generated by this motion of the charge − and no gravitational wave is generated by the geodesic motion of a point mass.

If we add some internal actions − as hydrodynamical pressures, *etc*. − the result does not change: indeed, in this case the world lines of the particles do not coincide with geodesic lines, but any acceleration of the new motion of a given point mass can be reproduced by a geodesic motion of the same particle in a suitable, "fixed", purely gravitational field.

A confirmation of the above argument is represented by the following remark. In the well-known Einstein-Infeld-Hoffmann method for deriving the motions of pointlike singularities of the gravitational field from Einstein equations, there is the possibility − at each stage of approximation − to pass to a reference frame for which our system of particles is *non*-radiating. And this is **not** an *ad hoc* trick: on the contrary, the mentioned possibility is rooted in the basic structure of general relativity. (Fock was of a different opinion, because he attributed a physically privileged role to the harmonic frames [4]. However, Fock's conviction − as emphasized by Einstein − is repugnant to the intrinsic nature of general relativity).

The conclusion is obvious: the undulatory solutions of Einstein field equations do not describe *physical* phenomena.